\documentclass{nature}[12pt]
\usepackage{amsmath}
\usepackage{amssymb}
\usepackage{color}
\usepackage{graphicx}
\usepackage{epstopdf}

\title{The Galaxy's Veil of Excited Hydrogen}

\author{Huanian Zhang$^{1}$\textsuperscript{*} \& Dennis Zaritsky$^1$}

\makeatletter
\let\saved@includegraphics\includegraphics
\AtBeginDocument{\let\includegraphics\saved@includegraphics}
\renewenvironment*{figure}{\@float{figure}}{\end@float}
\makeatother

\begin{document}

\maketitle

\begin{affiliations}
 \item Steward Observatory, University of Arizona, Tucson, AZ, USA
\end{affiliations}

{\bf \large \noindent Abstract}

{\bf 
Many of the baryons in our Galaxy probably lie outside
the well known disk and bulge components. Despite a wealth of evidence for the presence of some gas in galactic halos, including absorption line systems in the spectra of quasars, high velocity neutral hydrogen clouds in our Galaxy halo, line emitting ionised hydrogen originating from galactic winds in nearby starburst galaxies, and the X-ray coronas surrounding the most massive galaxies, accounting for the gas in the halo of any galaxy has been observationally challenging primarily because of its low density in the 
expansive halo. The most sensitive measurements come from detecting absorption by the intervening gas in the spectra of distant objects such as quasars or distant halo stars, but these have typically been limited to a few lines of sight to sufficiently bright objects. Massive spectroscopic surveys of millions of objects provide an alternative approach to the problem.
Here, we present
the first evidence for a widely distributed, neutral, excited hydrogen component of the Galaxy's halo. It is observed as the slight,  (0.779 $\pm$ 0.006)\%, absorption of flux near the rest wavelength of H$\alpha$ in the combined spectra of hundreds of thousands of galaxy spectra and is ubiquitous in high latitude lines of sight.
This observation provides an avenue to tracing, both spatially and kinematically, the majority of the gas in the halo of our Galaxy.}

Studies of the halo gas in either our Galaxy\cite{bregman}, or similar galaxies\cite{werk}, agree that the mass of the halo gas is comparable to that of the stars and cold disk gas within galaxy disks. 
Having roughly half the baryonic mass in the Galaxy's halo also reconciles measurements of the dynamical mass of the Galaxy and the cosmological baryon fraction \cite{zaritsky}. 
There are many previously identified halo gas components, but the component we describe
here is distinct from
the isolated clouds of neutral hydrogen\cite{wakker},  the ionised hydrogen 
associated with those clouds\cite{weiner}, and the cool ($T \sim 10^{4}$ K) and hot ($T \gtrsim 10^5$ K) diffuse components observed either in absorption in the spectra of select halo stars or as diffuse X-ray emission \cite{sembach,collins,kerp,murga,gupta,miller,nicastro} in that it is both pervasive and likely to be directly linked to the dominant mass component, cool, diffuse hydrogen. We present evidence that the absorption arises from halo gas with velocities that reach the Galactic escape speed, but which is on average neither rotating about the Galaxy nor rapidly falling inward or expanding outward from the Galaxy.
These absorption lines may ultimately provide the best diagnostics for measuring the kinematics, spatial distribution, and temperature structure of the dominant baryonic component of the Galaxy.

{\bf \large \noindent Results: Detecting a Signal}

While measuring the recombination radiation from ionised hydrogen in the halos
of other galaxies using millions of spectra\cite{zhang} from the 12th data release of the Sloan Digital Sky Survey (SDSS)\cite{sdss}, we 
observed H$\alpha$ absorption at rest, in the observed frame, not in individual spectra but in the average of 
thousands of spectra.
This finding motivated us to expand the sample of archival spectra that we examined beyond those that were suited to that earlier study and to
produce the mean continuum-normalised spectrum shown in Figure 1. The difference between the earlier and current samples is that we allow for spectra
with higher continua and exclude lines of sight with little or no continua in the observed wavelength range 6340 to 6790 \AA\ ($1\times10^{-17} \le {\rm mean\  flux}~({\rm ergs \ } {\rm cm}^{-2} {\rm s}^{-1} {\rm \AA}^{-1}) \le 5\times10^{-17}$) because detecting absorption requires a background illuminating source.

We stack all 732,225 galaxy spectra in our sample to obtain the highest possible signal-to-noise measurement of the 
H$\alpha$ absorption line profile (Figure 1). 
Because absorption is weak along each line of sight, we are in the regime where the stack represents
the linear superposition of millions of physical absorbing clouds. As such, it provides both a measure of the kinematic properties of the cloud ensemble
and a measure of the mean column density of hydrogen in the $n=2$ quantum state along a line of sight within the mapped region. At its strongest, the absorbing material removes on average
about 1\% of the incident flux at the wavelength of H$\alpha$. We present details of the procedure and discuss additional tests of the significance of the detection in the Methods section.

{\bf \large \noindent Results: The Gas Kinematics}

A striking feature of the stacked spectrum is the width of the H$\alpha$ line. When we measure the width at an absorption level of 
0.5\%, we find that it corresponds to line of sight velocities of approximately $\pm$ 700 km s$^{-1}$. 
Because we do not know the distribution of gas along the velocity axis, there is uncertainty in
determining the underlying, physical velocity spread. For example, if we assume that all of the gas is at extreme positive and negative velocities, those velocities need to be only  $\sim \pm$390 km s$^{-1}$, when convolved with the SDSS spectra resolution, to produce the measured width.
For comparison, if we assume that the gas is equally distributed in velocity between a lower and upper limiting velocity, then the inferred maximum velocities are $\sim \pm 680$ km s$^{-1}$. Although the inferred velocities are quite different in the two scenarios, in both cases they are 
inconsistent with the velocity expected for gas in the rotating Galactic disk. Because the gas does not have the characteristic of disk gas, some of the velocity range must be due to the Solar reflex motion.

A natural conjecture is 
to associate these maximum velocities with the Galaxy's escape velocity, which would surely place much of this gas in the halo. Estimates of the escape speed 
have been obtained using Galactic hypervelocity stars\cite{fragione}. 
For models with preferred Galactic masses, the escape speed can vary from 
550 to 650 km s$^{-1}$ at small Galactocentric distances ($\sim$ 10 kpc) to a few hundred km s$^{-1}$ at larger distances. As such, the
maximum velocities we measure are a plausible match to the escape velocity, although we need a model of how the gas is distributed throughout the halo to make a detailed quantitative
assessment. Velocities of this magnitude also match what is observed\cite{putman} for a known halo population,  high velocity H{\small I} clouds, $-500$ to $450$ km s$^{-1}$. 
We conclude that much of the absorption we observe comes from gas in the halo
of our galaxy. The exact nature of the gas, whether it is bound, infalling, being ejected, the
signs of a Galactic fountain, or something else await further modeling. An additional complication in the interpretation of the velocity distribution is that gas with a velocity greater than the escape speed will not necessarily escape the Galaxy because it is likely to interact with other halo gas components on its way out and dissipate some of its energy. As such, it could be at distances larger than 10 kpc, have this high a velocity, and not ultimately be lost from the Galaxy.

Additional features are visible in the stack around H$\alpha$. 
A second absorption feature in the stacked spectra is visible at $\sim$ 6496 \AA. We interpret this feature as the blend of absorption lines from various elements mainly Ca I, Fe I, Ni I and Ba I
that is also seen in late-type stellar spectra\cite{leborgne}. The narrowness of the line, relative to what we find for H$\alpha$, suggests
that this absorption does not arise entirely from the same gas as that responsible for the H$\alpha$ absorption. A third feature, this time in emission, is present at $\sim$ 6707 \AA. However, this feature does not show the systematic motion with Galactic coordinates that the other lines show and that we describe next, and so we suspect its origin is either terrestrial or instrumental.

The wavelengths of SDSS spectra are vacuum
wavelengths in the heliocentric frame. As such, we expect absorption lines due to the sum of halo gas along lines of sight to exhibit the reflex motion of the Sun around the Galactic Centre.  To test
this expectation, we divide our data into $20^\circ \times 35^\circ$ longitude, $l$, and latitude, $b$, bins, within which we stack the spectra and measure the flux-weighted centroid velocity of the 6496 \AA\ absorption lines.  For H$\alpha$, we use $20^\circ \times 90^\circ$ longitude and latitude bins to improve the precision of the flux-weighted velocity centroids. We fit for the Solar motion, which in our model consists of a circular orbit (the Local Standard of Rest, LSR, motion), and for a peculiar component separately to the velocities determined for the 6496 \AA\ and H$\alpha$ absorption lines. 
We 
calculate the projected reflex Solar motion along each line of sight and minimise the residuals relative to the observed velocities across the entire sample. We treat the rest frame wavelength of the 6496 \AA\ absorption line as a free parameter because we do not know its exact central wavelength. 

The resulting best fit
parameters using the 6496 \AA\ line are an LSR motion of $226 \pm 16$ km s$^{-1}$ toward $l = 90^\circ$ and a peculiar Solar motion of $23\pm9$ km s$^{-1}$ in a direction of $l =47^\circ \pm 24^\circ$ and $b = -8^\circ\pm36^\circ$. The best fit parameters using H$\alpha$ are an LSR motion of  $190 \pm 18$ km s$^{-1}$ toward $l = 90^\circ$ and a peculiar Solar motion of $25\pm 16$ km s$^{-1}$ in a direction of $l = -30^\circ\pm 29^\circ$ and $b =6^\circ \pm 42^\circ$.
The results from this very simple, straightforward analysis are broadly consistent (within $\sim 1\sigma$) with a standard value\cite{dehnen} of the Solar motion of 13.4 km s$^{-1}$ toward an $(l,b)$ of $(28^\circ, 32^\circ)$. 
The standard deviation of the data about the simple model, 38 km s$^{-1}$ or 32 km s$^{-1}$ for the 6496 \AA\ and H$\alpha$ lines respectively, is only modestly higher than an internal estimate of the uncertainty in our line centroid measurements of 24 km s$^{-1}$. The good agreement between our results based on a model assuming a net static halo and the published Solar motion based on local stellar kinematics suggests that the absorption does arise from halo gas and that this gas has no large net rotation or radial motion. We are not claiming to rule out the small net infall velocities inferred on other grounds and used to estimate the disk gas accretion rate\cite{putman}.
To approximately visualise the fit, we ``deproject" the observed line-of-sight velocities by dividing by 
$\cos b$, which is entirely correct only if the Sun's velocity is confined to the Galactic plane.  When we do that, we obtain the results shown in Figure 2.
Systematic deviations between the model and fit, such as those perhaps present at $l > 300^\circ$,  will be investigated in a more thorough treatment of the kinematics in 
a subsequent study. This now understood dependence in the observed radial velocities of the gas on Galactic longitude will account for some of the velocity range seen in stacked spectrum of Figure 1. A smaller velocity 
range for the gas would imply that the gas might still 
reach the local escape velocity but be at a somewhat larger radii.

{\bf \large \noindent Results: The Physical State of the Gas}

We now proceed to discuss inferences regarding the amount of gas responsible for the H$\alpha$ absorption. The integral of the stacked
spectrum corresponds to a lost flux of $(3.39 \pm 0.02) \times 10^{-18}$ ergs s$^{-2}$ cm$^{-2}$ in a mean spectrum whose continuum level is $(1.987 \pm 0.001) \times 10^{-17}$ erg s$^{-1}$ cm$^{-2}$ \AA$^{-1}$ at H$\alpha$. The equivalent width, $W$, of the feature is therefore $(0.170 \pm 0.001)$ \AA. 
$W$ is related to the column density\cite{Zhu,murga}, $N$, of absorbing matter, $ N = 1.13 \times 10^{20} \frac{W}{f\lambda^2} \  {\rm \AA \  cm}^{-2}$,
where $f$ is the oscillator strength for the $n=2 \rightarrow 3$ transition in hydrogen\cite{wiese} (0.64108),  and $\lambda$ is the wavelength of the line centre in \AA. 
The implied column density of neutral hydrogen in the $n =2$ state along the average line of sight in the mapped region is
$(7.34 \pm 0.04) \times 10^{11}$ cm$^{-2}$. Converting our measurement to a total hydrogen column mass involves a large, and uncertain, correction because the halo hydrogen is
highly ionised\cite{werk}. Instead, we advocate using our measured value of the column density as an additional constraint on detailed, physical models of the Galactic halo. There are studies\cite{barger,fox} of H$\alpha$ emission in specific halo environments, such as in the Magellanic Bridge and Stream, where the column density of H{\small I} gas is estimated to be $\log{N(HI)} = 20.7\pm 16.8$\cite{barger} and $\log{N(HI)} = 18.63\pm 0.08$\cite{fox}.
These values give an indication of the order of magnitude of the level population and ionisation corrections that are likely to be necessary.  

The Wisconsin H$\alpha$ Mapper (WHAM) team presented an all-sky H$\alpha$ emission map, which is the dataset that provides the most direct comparison to our observations. The emission signal can be as strong as $>$ 100 R (Rayleighs) or 5.6 $\times 10^{-16}$ ergs s$^{-2}$ cm$^{-2}$ arcsec$^{-2}$ around the Galactic Equator \cite{WHAM} but is much lower at the Galactic latitudes examined here. Although their maps show that certain regions within our survey area can have emission as high as $\sim$ 1, they conclude that there is a floor emission of $\sim$ 0.1 R over most of the sky on the angular scale of the survey. 
This level of emission will fill in the absorption line by about 17\%, and is therefore not dominant.
However, the contamination is likely to be even lower than this level because the emission, due to its dependence on the electron number density squared, will be highly clumpy and will therefore appear only in a small number of our lines of sight, but with fluxes much larger than the mean value quoted above. 

We are selecting lines of sight that contain a bright background source, which, as for IGM studies using QSO's, allows us to 
probe low column densities of intervening material.  A related study is that using {\sl HST} to probe low redshift galactic halos using QSO absorption lines\cite{werk}. We compare the two studies using a very crude, highly simplified model of the halo gas. By combining the {\sl HST} results\cite{werk} and our previous work\cite{zhang}, we inferred a halo gas temperature of 12,000 K (at $\sim$ 100 kpc). If we assume
a uniform temperature and density, the ratio of H atoms in the $n$ = 2 state relative to the ground state is $1.9 \times 10^{-6}$. Adopting an ionisation fraction of 99\% \cite{werk}, and a total column density of hydrogen of $10^{19}$ cm$^{-2}$, which is a plausible, if highly uncertain, value\cite{Boisse,Kaplan,Neeleman}, then we calculate that the column density of $n=2$ H atoms is $2 \times 10^{11}$ cm$^{-2}$, only a factor of $\sim 3$ lower than the estimate based on our measurement. The number of assumptions and simplifications illustrates, however, why one cannot take our result and 
infer meaningful constraints on the halo gas mass.

{\bf \large \noindent Results: The Spatial Distribution of the Gas}

Finally, we present a map of the absorption, Figure 3, that is constructed using 
the mean absorption in continuum-normalised galaxy spectra 
calculated within a wavelength window centred on rest frame H$\alpha$ that corresponds to velocity differences of $\pm$ 700 km s$^{-1}$ along lines of sight 20$^\circ$ above and below the Galactic Plane. 
Each 5$^\circ$ by 5$^\circ$ angular cell on the sky inside of which we calculate the mean absorption typically encompasses several hundred to thousands of spectra. The map is a rebinned,  interpolated, and smoothed (on a scale of $\sim$ 4$^\circ$) version of
those mean values projected onto the celestial sphere represented in Galactic coordinates. 
This map represents the sum of emission and absorption of H$\alpha$ along
the entire line of sight through the Galaxy's halo. We have no direct constraint on the distance of the absorbing matter or even whether it is
primarily in a single cloud or distributed along the line of sight. Absorption is prevalent, without any highly distinct  large-scale structures. Despite the apparent homogeneity of the absorption in the Figure, we do find a marginally significant anti-correlation between the H$\alpha$ absorption depth and Galactic latitude (using a Spearman rank correlation analysis we estimate the probability of it happening randomly at only 1.3\%), suggesting that the distribution of absorbing material is somewhat flattened.

{\bf \large \noindent Conclusions}

In summary, we trace for the first time a distinct component of the Galactic halo, diffuse neutral hydrogen, using the small fraction of that hydrogen that in the $n=2$ state. We present 1) an integrated spectrum that suggests that the absorbing gas spans velocities up to the Galactic escape speed, 2) the measurement of an apparent kinematic signature in the gas that we associate with the Solar reflex motion, indicating that the gas is in the net mostly static in the Galactic frame, 3) the measurement of the equivalent width of the absorption feature for a typical line of sight at high Galactic latitudes from which we estimate a column density of the absorbing gas, and 4) a map of this halo component projected on the sky. Each of these observations, in their own way, support our conclusion that this gas is part of the Galactic halo. This component is distinct from
the previously identified isolated clouds of neutral hydrogen\cite{wakker}, ionised hydrogen 
associated with those clouds\cite{weiner}, and cool ($T \sim 10^{4}$ K) and hot ($T \gtrsim 10^5$ K) diffuse components\cite{sembach,collins,kerp,murga,gupta,miller,nicastro}.

The majority of the baryonic matter in our Galaxy lies outside
of the well known disk and bulge components\cite{bregman}.
The principal impact of the discovery of the component described here is how it will enable us to explore the elusive, but critical, baryonic halo. By measuring the H$\alpha$ absorption line profiles in the ever increasing number of lines of sight with spectroscopy, the kinematic modeling will no longer need to invoke a static halo but can instead explore and constrain more complex dynamical models. Such work will constrain the net flows of gas in, out, and around our halo. By measuring line absorption to halo stars at different distances it will become possible to map the 3D distribution of the halo gas. By measuring the absorption by other spectral features, such as H$\beta$, we will constrain the temperature distribution of the gas, or using other features the elemental abundances in the halo gas. The full set of absorption features from this component may provide the best approach to understanding the nature of the dominant baryonic component of the Galaxy.

{\bf \large \noindent References}
\bibliographystyle{naturemag}
\bibliography{paper}

\begin{addendum}
 \item DZ and HZ acknowledge financial support from NASA ADAP NNX12AE27G, NSF AST-1311326, and the University of Arizona. The authors thank the SDSS III team for providing a valuable resource to the community. Funding for SDSS-III has been provided by the Alfred P. Sloan Foundation, the Participating Institutions, the National Science Foundation, and the U.S. Department of Energy Office of Science. The SDSS-III web site is http://www.sdss3.org/.
SDSS-III is managed by the Astrophysical Research Consortium for the Participating Institutions of the SDSS-III Collaboration including the University of Arizona, the Brazilian Participation Group, Brookhaven National Laboratory, Carnegie Mellon University, University of Florida, the French Participation Group, the German Participation Group, Harvard University, the Instituto de Astrofisica de Canarias, the Michigan State/Notre Dame/JINA Participation Group, Johns Hopkins University, Lawrence Berkeley National Laboratory, Max Planck Institute for Astrophysics, Max Planck Institute for Extraterrestrial Physics, New Mexico State University, New York University, Ohio State University, Pennsylvania State University, University of Portsmouth, Princeton University, the Spanish Participation Group, University of Tokyo, University of Utah, Vanderbilt University, University of Virginia, University of Washington, and Yale University.
 \item[Author contributions] Both authors contributed to the final analysis and interpretation of the results. Huanian Zhang lead the data analysis. Dennis Zaritsky provided the initial motivation for the program.
\item[Author Affiliation]  Steward Observatory, University of Arizona, Tucson, AZ, USA
\item[Competing financial Interests] The authors declare that they have no
competing financial interests.
 \item[Materials \& Correspondence] Correspondence and requests for materials
should be addressed to Huanian Zhang ~(email: fantasyzhn@email.arizona.edu).
\end{addendum}

\begin{figure}
\begin{center}
\includegraphics[width = 0.8 \textwidth]{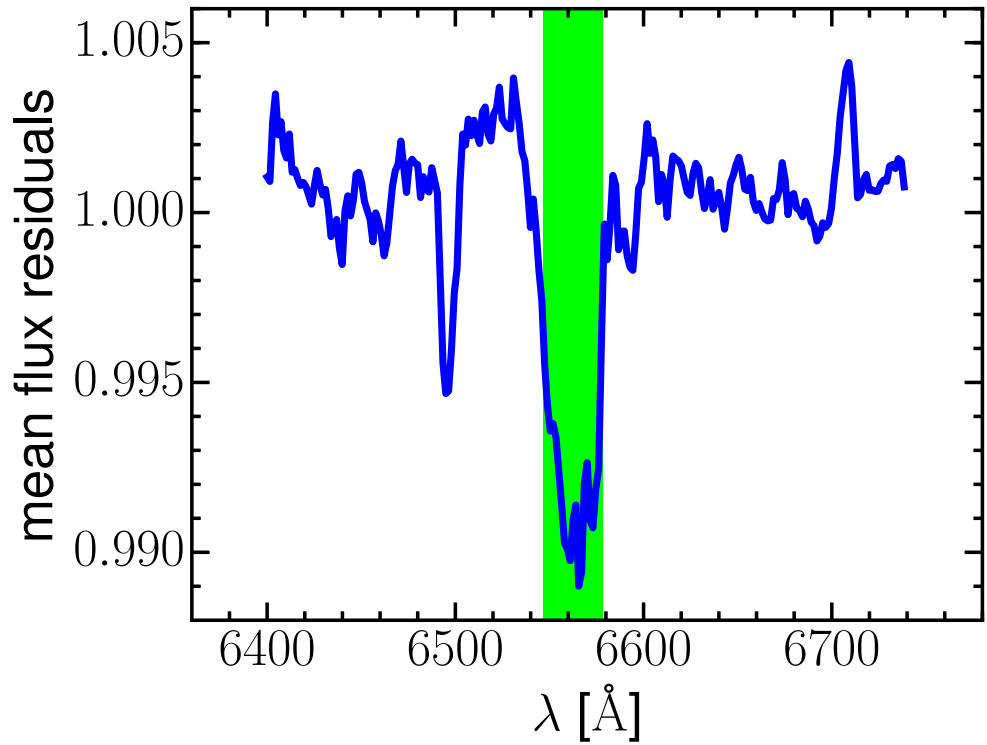}
\end{center}
\caption{The combined, continuum-normalised, line of sight spectrum in the region of rest frame H$\alpha$ for over seven hundred thousand SDSS galaxy spectra. The shaded region corresponds to a velocity spread of $\pm~700$ km s$^{-1}$.}
\label{fig:stackall}
\end{figure}

\begin{figure}
\begin{center}
\includegraphics[width = 0.8 \textwidth]{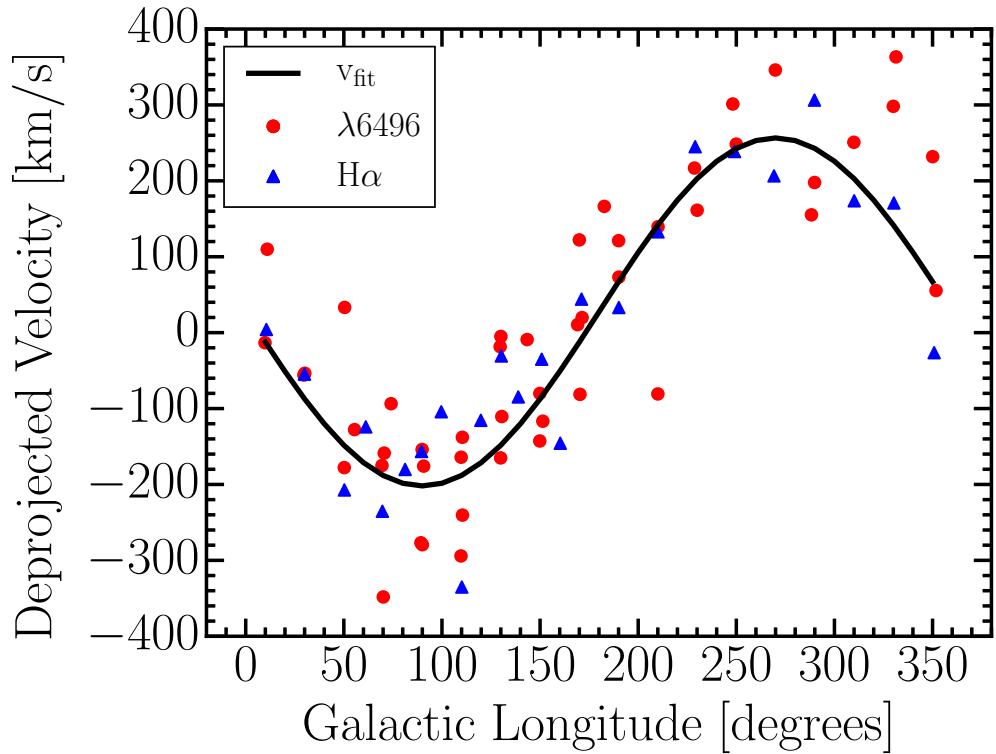}
\end{center}
\caption{The apparent motion of the absorbing gas as a function of longitude as measured from the H$\alpha$ and 6496 \AA\ absorption lines. The black line is the expected reflex motion of the Sun. The parameters that describe that motion come from a least-square fit to the $\lambda$6496 data with parameters given in the text, which agree with independent determinations of the Solar motion.
The  agreement between the data and simple model suggests that the gas is nearly at rest (in the net) within the Galaxy, consistent with it being a halo component. The plotted points correspond to the bins described in the text.}
\label{fig:motion}
\end{figure}

\begin{figure}
\begin{center}
\includegraphics[width = 0.8 \textwidth]{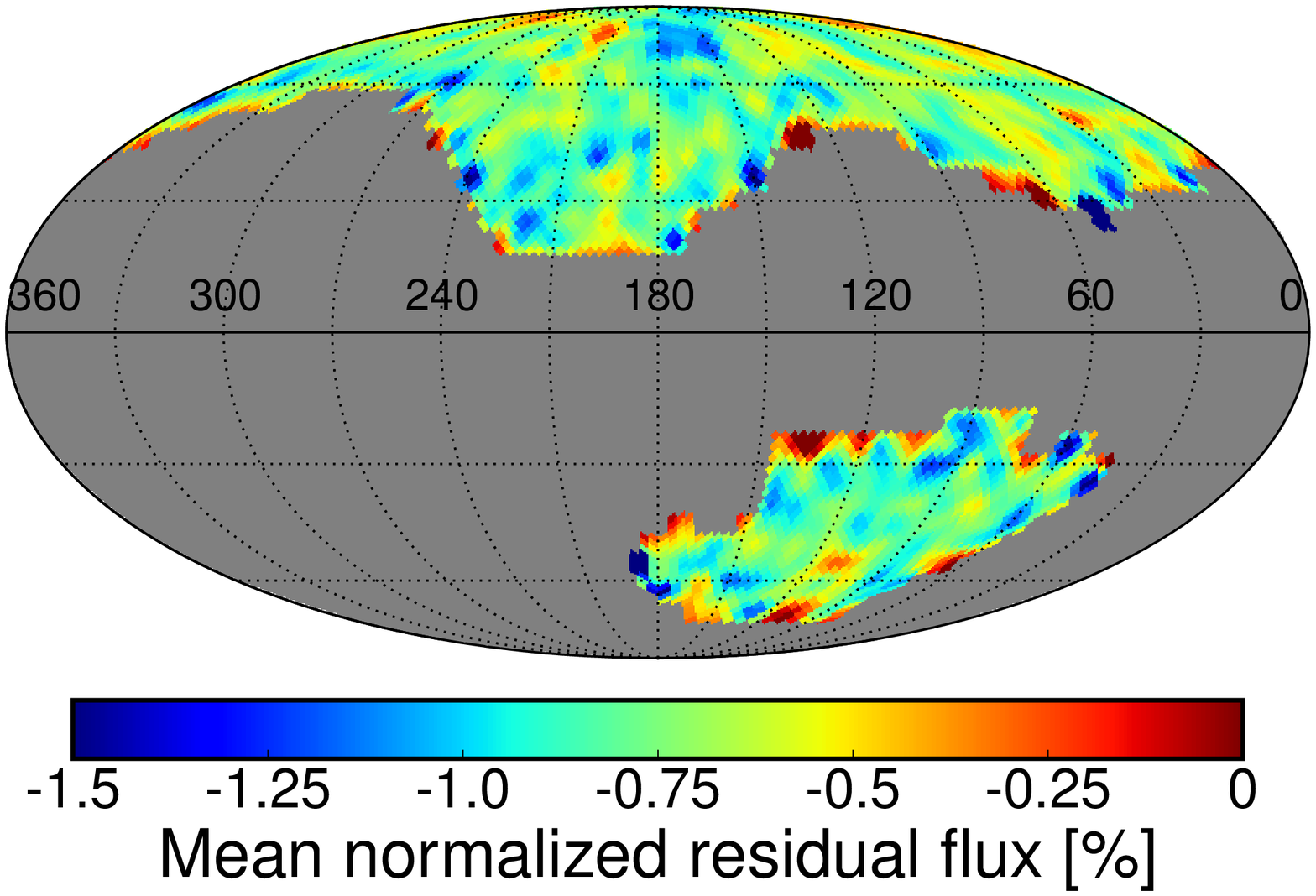}
\end{center}
\caption{The mean H$\alpha$ absorption map calculated within a wavelength window centred on rest frame H$\alpha$ that extends $\pm~700$ km s$^{-1}$ in the SDSS continuum-normalised galaxy spectra. We present an interpolated version of our measurements projected on the celestial sphere using the Hammer projection.  To visually enhance the contrast, a limited number of regions are off-scale. Galactic longitudes are labeled. Galactic Latitude varies from $+90^\circ$ at the top to $-90^\circ$ at the bottom. }
\label{fig:map}
\end{figure}

\begin{methods}

The SDSS spectra are wavelength-calibrated, flux-calibrated, and sky-subtracted with a resolution of 1500 at 3800 \AA, 2500 at 9000 \AA\cite{2012AJ....144..144B}. We remove the continuum as done in our previous study\cite{zhang}. We measure the H$\alpha$ flux (or decrement) in a window corresponding to $\pm$ 700 km s$^{-1}$. We remove 2$\sigma$ outliers from the distribution of values so that the stack is not dominated by outliers.
We stack the individual spectra for the wavelength region of 6400 \AA\ to 6730 \AA\ in the observed frame, pixel by pixel, with equal weight.

We consider our detection and interpretation valid after the following tests and considerations. First, we estimate the uncertainty of the flux decrement using the empirically determined dispersion among all of the measured values. The mean absorption value and its uncertainty are $0.779\pm0.006$\%. The internal error suggests that the detection is highly significant.
Second, we divide the entire data into ten subsets, leaving one out each time when we do the stack. The results are consistent. 
The mean absorption values range from 0.772\% to 0.789\%, with uncertainties that are $0.006$\%, so the extremes differ from the mean value by about 1.5$\sigma$. 
Third, we produce stacks using background galaxies of different mean continua levels (within the specified limits).  If the absorption originates from contamination (e.g.,  halo stars projected onto the line of sight), we would expect the signal to decrease in proportion to the increasing background source signal.  We have divided the data into nine bins by continuum flux. The H$\alpha$ absorption values in the standard window vary in the range of 0.63\% to 0.92\%, but do not 
correlate with continuum value (a Spearman rank correlation analysis results in a probability of a deviation from random of 0.24, so not significant, and the trend, if any, is in the opposite sense as that expected in this scenario). Lastly, our results relating the detection (either absorption strength or kinematics) to Galactic coordinates demonstrate that the features are not terrestrial or instrumental. 

Because we remove spectra
that have anomalous flux levels in the H$\alpha$ window (either high or low) from our stack, the lines of sight with the strongest emission will have been removed. 
The angular resolution of their survey is 1$^\circ$, so we cannot do a direct comparison of our spectra and their results. 
A caveat to any interpretation our absorption line profile is that the emission that does remain in our stack is likely, according to the WHAM results, to be confined to a narrow velocity window ($\sim 100$ km s$^{-1}$) and so will preferentially fill near the line centre. As such, there is likely some distortion by line filling of the overall profile shape presented in Figure 1. 

{\bf Data Availability Statement}

The data that support the plots within this paper and other findings of this study are available from the corresponding author upon reasonable request. The SDSS spectra can be obtained through http://www.sdss.org/dr12.


\end{methods}




\end{document}